%% file: gdm.tex
\documentclass[11pt, a4paper, copyright, gdm]{google}

\usepackage[authoryear, sort&compress, round]{natbib}
\input{setup/packages}
\input{setup/custom_commands}
\input{defines}


\uselogo{} 

\title{Designing Proactive Thought Partners for Writing}

\correspondingauthor{Chao Zhang, cz468@cornell.edu}

\reportnumber{} 

\renewcommand{\today}{}

\author[*, 1]{Chao Zhang}
\author[1]{Abe Davis}
\author[2]{Chih-Wei Chen}
\author[2]{Chin-Chia Hsu}

\affil[*]{Work done during an internship at Google DeepMind.}
\affil[1]{Cornell University}
\affil[2]{\thepa{}{}}

\begin{abstract}
\input{sections/0-abstract}
\end{abstract}

\begin{document}

\maketitle

\input{sections/1-introduction-v2}
\input{sections/2-related-work}
\input{sections/4-system-design}
\input{sections/5-evaluation}

\input{sections/6-results}
\input{sections/7-discussion}
\input{sections/8-conclusion}

\bibliography{reference}


\end{document}

%% file: setup/packages.tex
\usepackage{acmart-taps}

\usepackage{tabularray}
\usepackage{multirow} 
\usepackage{array} 
\newcolumntype{C}[1]{>{\centering\arraybackslash}p{#1}} 

\usepackage{wrapfig} 
\usepackage{float} 

\usepackage{siunitx} 
\usepackage{etoolbox} 
\usepackage{dcolumn} 
\usepackage{booktabs} 

\usepackage{fontawesome5}

\usepackage{xcolor} 
\usepackage{soul} 
\soulregister{\cite}7 
\soulregister{\citep}7
\soulregister{\citet}7
\soulregister{\ref}7
\soulregister{\pageref}7

\usepackage[most]{tcolorbox}

\newtcolorbox{myfancybox}{
  colback=light-bg,
  colframe=black,
  boxrule=0.8pt,
  arc=0pt,
  outer arc=0pt,
  left=6pt,
  right=6pt,
  top=6pt,
  bottom=6pt,
  before skip=8pt,
  after skip=8pt,
  parskip=3pt
}

\usepackage{xspace} 

\usepackage{bm} 
\newrobustcmd*{\bftabnum}{ %
  \bfseries
  \sisetup{output-decimal-marker={\textmd{.}}} %
}

%% file: setup/custom_commands.tex
\definecolor{oxfordblue}{rgb}{0.0, 0.13, 0.28}
\definecolor{harvardcrimson}{rgb}{0.79, 0.0, 0.09}
\definecolor{dartmouthgreen}{rgb}{0.05, 0.5, 0.06}
\definecolor{princetonorange}{rgb}{1.0, 0.56, 0.0}
\definecolor{yaleblue}{rgb}{0.06, 0.3, 0.57}
\definecolor{usccardinal}{rgb}{0.6, 0.0, 0.0}
\definecolor{uclablue}{rgb}{0.33, 0.41, 0.58}
\definecolor{msugreen}{rgb}{0.09, 0.27, 0.23}
\definecolor{cornellred}{rgb}{0.7, 0.11, 0.11}
\definecolor{pomegranate}{RGB}{192, 57, 43}
\definecolor{anti-pomegranate}{RGB}{43,178,192}
\definecolor{alizarin}{RGB}{231, 76, 60}
\definecolor{anti-belize}{RGB}{185, 41, 56}
\definecolor{belize}{RGB}{41, 128, 185}
\definecolor{sky}{RGB}{52, 152, 219}
\definecolor{green}{RGB}{22, 160, 133}
\definecolor{anti-green}{RGB}{160,22,118}
\definecolor{turquoise}{RGB}{26, 188, 156}
\definecolor{pumpkin}{RGB}{211, 84, 0}
\definecolor{anti-pumpkin}{RGB}{0,22,211}
\definecolor{carrot}{RGB}{230, 126, 34}
\definecolor{wisteria}{RGB}{142, 68, 173}
\definecolor{anti-wisteria}{RGB}{99,173,68}
\definecolor{amethyst}{RGB}{155, 89, 182}
\definecolor{nephritis}{RGB}{39, 174, 96}
\definecolor{anti-nephritis}{RGB}{174,39,117}
\definecolor{grey-bg}{RGB}{242,242,235}
\definecolor{light-bg}{RGB}{249,249,249}
\definecolor{extended-blue}{RGB}{59,130,246}
\definecolor{extended-red}{RGB}{239,68,68}
\definecolor{extended-orange}{RGB}{249,115,22}
\definecolor{extended-violet}{RGB}{99,102,241}
\definecolor{extended-green}{RGB}{16,185,129}

\newcommand{\eg}{e.g.,\ }

\newcommand{\ie}{i.e.,\ }

\AtBeginDocument{ %
  \providecommand\BibTeX{{ %
    \normalfont B\kern-0.5em{\scshape i\kern-0.25em b}\kern-0.8em\TeX}}}

\newcommand{\namedparagraph}[1]{\vspace{0.2cm}\noindent\textbf{#1:}}

\newcommand{\formatcaption}[2]{\textit{#1} \textmd{#2}}

\newcommand{\fig}[1]{Fig. {#1}}
\newcommand{\figref}[1]{\fig{\ref{#1}}}
\newcommand{\tab}[1]{Table {#1}}
\newcommand{\tabref}[1]{\tab{\ref{#1}}}
\newcommand{\secref}[1]{§\ref{#1}}

\usepackage{enumitem}
\newenvironment{packeditemize}{
\begin{itemize}[leftmargin=0.5cm]
\setlength{\itemsep}{1pt}
\setlength{\parskip}{2pt}
\setlength{\parsep}{0pt}
}{\end{itemize}}

\usepackage{pifont}  

\newcommand{\taglabel}[1]{\textsf{\scriptsize\textbf{[#1]}}}

\usepackage[most]{tcolorbox}
\tcbset {
  base/.style={
    arc=0mm, 
    bottomtitle=0.5mm,
    boxrule=0mm,
    colback=black!2!white,
    colbacktitle=black!5!white,
    coltitle=black, 
    fonttitle=\bfseries, 
    left=2.5mm,
    leftrule=1mm,
    right=3.5mm,
    title={#1},
    toptitle=0.75mm, 
  }
}

\definecolor{brandblue}{rgb}{0.34, 0.7, 1}
\newtcolorbox{takeawaybox}[1]{
  colframe=brandblue, 
  base={#1}
}

\newtcolorbox{subbox}[1]{
  colframe=black!30!white,
  base={#1}
}

%% file: defines.tex
\newif\ifsubmit
\submitfalse

\input{macros/NotesAndEdits}


%% file: macros/NotesAndEdits.tex
\ifsubmit
\definecolor{author_colorA}{rgb}{0,0.5,1}
\definecolor{author_colorB}{rgb}{0.2,.64,0}
\definecolor{author_colorC}{rgb}{1,0,1}
\definecolor{author_colorD}{rgb}{0,1,1}
\definecolor{changes_color}{rgb}{0.05,0.5,0.3}
\definecolor{mathbrace_color}{rgb}{0.2,0.5,1.0}
\fi

\ifsubmit
    
    \newcommand{\abe}[1]{}
    \newcommand{\authorB}[1]{}
    \newcommand{\authorC}[1]{}
    \newcommand{\authorC}[1]{}
    
    \newcommand{\URGENT}[1]{}
\else
    
    \newcommand{\abe}[1]{\textbf{\textcolor{blue}{AD: #1}}}
    \newcommand{\authorB}[1]{\textbf{\textcolor{author_colorB}{CS: #1}}}
    \newcommand{\authorC}[1]{\textsf{\textcolor{author_colorC}{[{\bf AUTHORB}: #1]}}}

    \newcommand{\URGENT}[1]{{\textcolor{red}{URGENT:[#1]}}}
\fi

%% file: sections/0-abstract.tex
Writing involves diverse cognitive activities, from ideation to revision, and writers' needs vary across individuals and moments. Proactive AI promises to provide the right support at the right time, yet existing proactive tools largely focus on generic textual assistance, such as autocomplete.
This paper studies the design space of \textit{proactive thought partners}: AI agents that proactively offer customizable, higher-level cognitive support during writing.
We instantiated this concept in a technology probe and deployed it with 16 participants for one week.
The probe allows users to create partners by configuring their roles and proactivity.
As users write, relevant partners take the initiative at appropriate moments to offer suggestions.
Our findings show that participants configured proactive support through prospective planning, used suggestions for both idea generation and self-monitoring, and valued lightweight visual representations alongside non-directive rhetorical framing for non-intrusive interventions.
We derive implications for designing proactive writing assistants around customization, timing, engagement, and representation.\looseness=-1

%% file: sections/1-introduction-v2.tex
\section{Introduction}

Writing is a dynamic cognitive process~\citep{flowerCognitiveProcessTheory1981a}.
From one moment to the next, a writer might shift between developing ideas, selecting prose, and evaluating written text in light of broader narrative goals.
The support they need can therefore change throughout the writing process.
At one moment, a writer may benefit from an example that makes an abstract claim concrete; at another, they may need help considering an alternative perspective or deciding how to develop the next section.
These needs also vary across writers.
Some writers may want support that helps them develop and evaluate their own ideas, while others may prefer information-seeking assistance, such as suggesting evidence from outside their domains.
An ideal writing assistant should therefore support this shifting and personal cognitive work by \textit{offering the right kind of help at the right moment}.\looseness=-1

Existing proactive writing assistance approaches this ideal in a narrower setting.
At the level of local text production, writing is relatively structured and predictable: the surrounding text often provides strong signals about useful continuations, while observable interaction events, such as pauses, can provide natural opportunities for intervention~\citep{chen_gmail_2019}.
Prior work has therefore leveraged such signals to proactively provide local textual assistance, such as autocomplete and next-phrase suggestions~\citep{bhatInteractingNextPhraseSuggestions2023,buschekImpactMultipleParallel2021a,jakeschCoWritingOpinionatedLanguage2023a}.
However, writing involves more than choosing words.
Higher-level cognitive activities such as ideation, reflection, and revision are more dependent on the writer's preferences and evolving context~\citep{flowerCognitiveProcessTheory1981a}.
In contrast to local textual assistance, support for these higher-level activities has largely relied on reactive interfaces, where tools respond only after users explicitly request assistance through prompts, buttons, or menus~\citep{geroSparksInspirationScience2022,zhangFrictionDecipheringWriting2025,rezaABScribeRapidExploration2023}.

\input{figures/fig-designspace}

We study proactive AI assistants that support these higher-level, personal cognitive needs in writing, which we call \textbf{\textit{proactive thought partners}} (\figref{fig:design_space}).
We instantiated this concept in a technology probe~\citep{hutchinsonTechnologyProbesInspiring2003}, using customization as both an interaction design mechanism and an inquiry instrument for exploring the design space.
The probe allows users to create thought partners by customizing both their \textit{roles} and \textit{proactivity}.
First, users define each partner's role to specify the type of cognitive support they need.
Second, users define timing conditions to specify when the partner should intervene. As users write, relevant partners take the initiative at appropriate moments to offer suggestions, which users can easily dismiss, use as inspiration, or ask the partner to apply directly to the document.

We deployed this probe with 16 participants in a one-week diary study followed by semi-structured interviews.
Our findings show how users customized, engaged with, and experienced this new form of proactive support for writing.
Participants used customization to translate anticipated needs and likely difficulties into different forms of cognitive support and timing conditions, treating configuration as an act of prospective planning.
During writing, they often ignored interventions to preserve flow, while engaged suggestions supported both idea generation and self-monitoring.
Participants also found proactive support less intrusive when it was presented through lightweight visual representations and non-directive rhetorical framing.
Our findings suggest design implications for proactive thought partners across four dimensions of the design space: customization, timing, engagement, and representation.
\looseness=-1

In this work, we contribute:

\begin{packeditemize}
    \item \textbf{Proactive thought partners}, a new design concept for intelligent writing assistants that proactively provide higher-level and personal cognitive support through customizable AI partners.
    \item \textbf{A technology probe} that enables writers to create proactive thought partners by configuring their roles and proactivity, allowing us to study customizable, proactive cognitive support in situated writing practice.
    \item \textbf{Empirical insights and design implications} from a one-week diary study with 16 participants, revealing how users customize proactive cognitive support to suit their needs and how they engage with proactive suggestions during writing. These findings articulate a design space for proactive thought partners in writing.
\end{packeditemize}

%% file: figures/fig-designspace.tex
\begin{figure*}
\centering
\includegraphics[width=\linewidth]{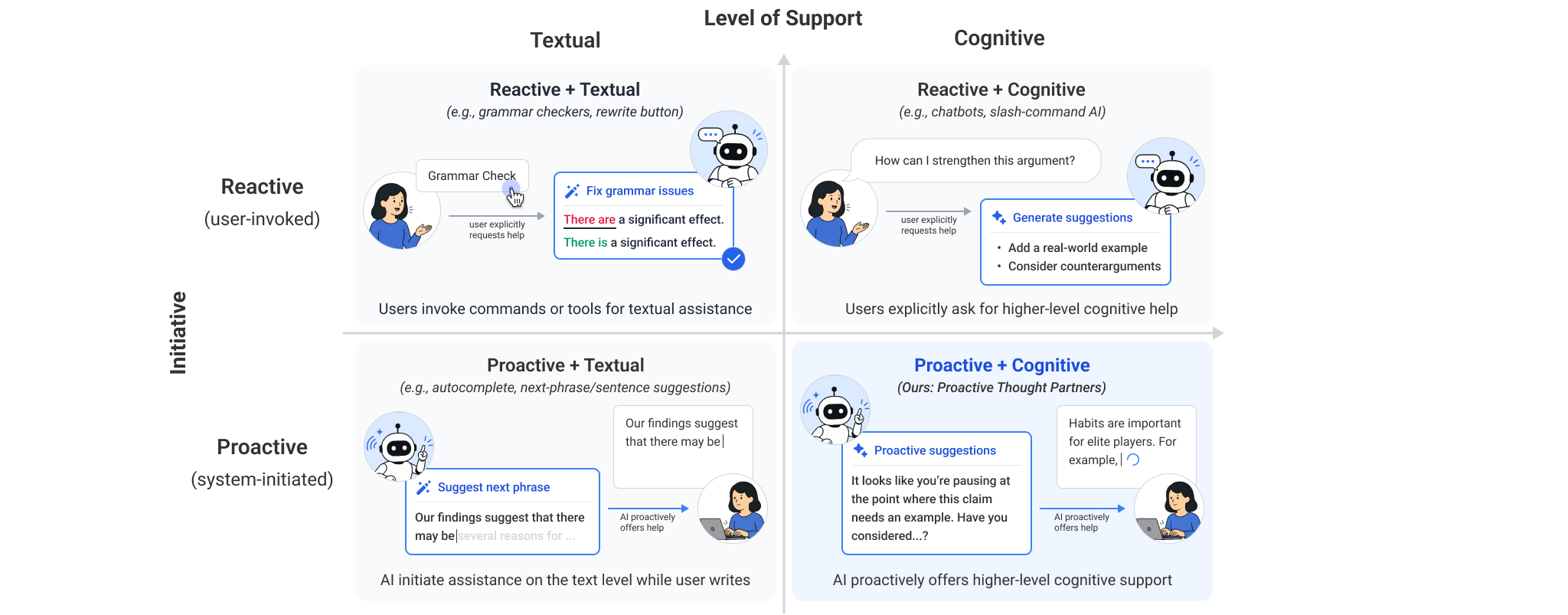}
\caption{\formatcaption{Proactive thought partners in the design space of intelligent writing assistants.}{We position intelligent writing assistants along two dimensions: initiative, distinguishing reactive user-invoked support from proactive system-initiated support, and level of support, distinguishing textual assistance from higher-level cognitive assistance. Proactive thought partners fall within the proactive and cognitive quadrant, where AI offers higher-level cognitive support during writing without requiring explicit user prompting.}}
\label{fig:design_space}
\end{figure*}

%% file: sections/2-related-work.tex

\section{Related Work}
In this section, we first situate our work within the broader landscape of intelligent writing tools.
We then review prior research on proactive AI assistants and customizable AI agents to discuss how our work extends these lines of research.

\subsection{Intelligent Writing Tools}
Flower and Hayes's cognitive process theory of writing understands writing as a recursive cognitive process involving planning, translating ideas into text, and reviewing what has been written~\citep{flowerCognitiveProcessTheory1981a}.
Intelligent writing tools have long supported parts of this process by helping writers produce, correct, or refine text~\citep{leeDesignSpaceIntelligent2024a}.
For example, autocomplete and predictive text systems offer word- or phrase-level continuations~\citep{bhatInteractingNextPhraseSuggestions2023,buschekImpactMultipleParallel2021a,jakeschCoWritingOpinionatedLanguage2023a,arnoldPredictiveTextEncourages2020}, while spelling, grammar, and style checkers~\citep{changWriteAhead2MiningLexical2015a,chungReadinghelpSupportingEFL2025,grammarlyGrammarlyFreeAI2025} help identify local errors or improve surface-level expression.
These systems can reduce the effort of text production and revision by providing relatively narrow forms of (sometimes proactive) assistance focused on the immediate text.\looseness=-1

Recent advances in large language models (LLMs) have expanded the scope of AI writing support.
Rather than only correcting or completing text, LLM-based systems can help writers brainstorm ideas~\citep{geroSparksInspirationScience2022,huangHeteroglossiaInSituStory2020,chouTaleStreamSupportingStory2023}, generate outlines~\citep{wanItFeltHaving2023,wanPolymindParallelVisual2025}, draft passages~\citep{zhangWordsWidgetsControllable2026a,zhangNarrativeKeyframingGenerative2026,zhangNarrixRemixingNarrative2026}, rewrite text~\citep{massonTextoshopInteractionsInspired2025,itoUseAIpoweredRewriting2023,leeInteractiveChildrensStory2022}, provide feedback~\citep{liuCraftingTextCrafting2026,benharrakWriterDefinedAIPersonas2024a,zhangFrictionDecipheringWriting2025}, and explore alternatives~\citep{rezaABScribeRapidExploration2023,zhangSynthiaVisuallyInterpreting2025} across a wide range of writing contexts~\citep{zhangVISARHumanAIArgumentative2023,zhangFrictionDecipheringWriting2025,chungTaleBrushSketchingStories2022,yuanWordcraftStoryWriting2022,huangHeteroglossiaInSituStory2020,shenConvXAIDeliveringHeterogeneous2023a,sunMetaWriterExploringPotential2024}.
This line of work shows that AI writing assistants can support higher-level cognitive activities such as ideation~\citep{geroSparksInspirationScience2022,huangHeteroglossiaInSituStory2020}, reflection~\citep{zhangFrictionDecipheringWriting2025}, and revision~\citep{zhangSynthiaVisuallyInterpreting2025} than accelerating text production.
However, much of this richer cognitive support remains reactive.
Writers often need to leave the flow of writing to compose prompts, select text, click buttons, or otherwise explicitly invoke the system.
This places the burden on writers to recognize when they need help, formulate the right request, and decide which kind of support to seek.
At the same time, more proactive writing support has largely been associated with predictive text, grammar diagnosis, and next-phrase suggestions~\citep{bhatInteractingNextPhraseSuggestions2023,buschekImpactMultipleParallel2021a,jakeschCoWritingOpinionatedLanguage2023a,arnoldPredictiveTextEncourages2020,changWriteAhead2MiningLexical2015a,tsaiLinggleWriteCoachingSystem2020}.
This creates a gap between two forms of assistance: systems that are proactive but focused on text continuation/correction, and systems that provide broader cognitive support but require explicit user initiation.
Our work explores the design space in this gap that combines proactivity and cognitive support in writing, which we term ``proactive thought partners.''\looseness=-1


\subsection{Proactive AI Assistants}

Rather than waiting for users to formulate a prompt or click a menu, proactive AI assistants take the initiative by offering assistance when they anticipate it may be useful.
This idea builds on a long tradition of mixed-initiative interaction, where both users and systems can take initiative in pursuit of a shared task~\citep{horvitzPrinciplesMixedinitiativeUser1999a}.
Recent work has explored proactive AI in domains such as programming~\citep{chenNeedHelpDesigning2025a,puAssistanceDisruptionExploring2025a,kuoDeveloperInteractionPatterns2026}, meetings~\citep{chenAreWeTrack2025}, design~\citep{sonWhenHandWhen2026}, surveys~\citep{liuSensingWhatSurveys2026}, tool manipulation~\citep{prasongpongchaiTalkHandLLMpowered2025}, scholarly recommendation~\citep{siangliulueOmakaseProactiveAssistance2026}, storytelling games~\citep{kreminskiWhyAreWe2020a}, web search~\citep{huangFacilitatingProactiveReactive2026}, problem-solving~\citep{luoHowUsersPerceive2026}, and group collaboration~\citep{mukhopadhyayExploringImpactProactive2026,zhouExploringNeedsDesign2026}, where systems may surface relevant information, reflect on goals, or suggest next steps.
In line with the most recent work on proactive AI~\citep{puAssistanceDisruptionExploring2025a,chenNeedHelpDesigning2025a,chenAreWeTrack2025} and an early Wizard-of-Oz study simulating proactive AI for writing~\citep{yinProactiveAICatalyst2026}, we define proactivity as the system actively inferring user needs and offering suggestions without requiring explicit user prompting.
\looseness=-1

Timing is key to system initiative~\citep{horvitzPrinciplesMixedinitiativeUser1999a,baileyEffectsInterruptionsTask2001}.
Existing proactive writing assistants often tie the timing of interventions to observable interaction events.
For example, autocomplete systems surface continuations as writers type~\citep{bhatInteractingNextPhraseSuggestions2023,buschekImpactMultipleParallel2021a,jakeschCoWritingOpinionatedLanguage2023a,arnoldPredictiveTextEncourages2020,chen_gmail_2019}, while~\citet{lehmannCollaborativeDocumentEditing2026} allowed users to bundle autonomous tasks with triggers such as pauses, edits, or saves.
In these systems, such events directly cue system action.
However, the same event can reflect different writing states.
A pause, for example, may indicate concentration, being stuck for ideas, or reviewing recently written text.
For higher-level cognitive support, interaction events alone may therefore be insufficient to determine whether assistance is needed and what kind of support would be useful.
Our work extends this event-based approach by introducing \textit{contextual heuristics}, which specify more specific timing conditions under which a partner should intervene when a broader interaction event occurs.
This distinction also allows us to study how users conceptualize these two forms of timing differently: event triggers as broad opportunities for the system to check whether support may be relevant, and contextual heuristics as more specific conditions for determining whether and which support should actually be provided.
In doing so, we examine how users expect proactive cognitive support to interpret their evolving writing context when deciding whether and which support to provide.

More closely aligned with our vision,~\citet{yinProactiveAICatalyst2026} conducted a Wizard-of-Oz study in which a human simulated proactive AI by interpreting the writing context and offering suggestions while another human wrote a story.
The study provides quantitative evidence that proactive AI, as simulated by a human, can enhance creativity by serving as a catalyst for inspiration rather than as a direct content provider.
However, it remains unclear how such proactive cognitive support should be designed in a functional system, how writing context should be interpreted to determine when support is appropriate, and how users might customize proactive thought partners to suit their own needs.
In this paper, we instantiate proactive cognitive support in a functional system with customizable thought partners and deploy it in a diary study to investigate how users configure, experience, and evaluate such support in practice.


\subsection{Customizable AI Agents}
A growing body of HCI research has explored customizable AI agents as a way to better align LLM-based systems with users' goals, preferences, and contexts.
Prior work has examined customization in domains such as writing~\citep{lehmannCollaborativeDocumentEditing2026,benharrakWriterDefinedAIPersonas2024a}, emotional support~\citep{zhengCustomizingEmotionalSupport2025}, social chatbots~\citep{haCloChatUnderstandingHow2024b}, research ideation~\citep{liuPerspectraChoosingYour2026}, and collaborative brainstorming~\citep{quanAIColleaguesMultiAgent2026}.
Across these settings, customization allows users to shape how an AI system should behave, what role it should take, and how it should respond to domain-specific needs.
For emerging AI interaction concepts, customization can also serve as a probe mechanism in a study~\citep{zhengCustomizingEmotionalSupport2025}: by observing what users choose to customize, researchers can surface latent needs, preferences, and expectations that may be difficult to anticipate a priori.

Customization is particularly important for writing because useful support depends on the writer's goals, genre, audience, voice, and stage in the writing process.
Given so, prior work has explored writer-defined AI personas for on-demand feedback~\citep{benharrakWriterDefinedAIPersonas2024a}, multiple user-defined agents and tasks in collaborative document editing~\citep{lehmannCollaborativeDocumentEditing2026}, and layered interfaces for aligning content development with rhetorical strategy~\citep{siddiquiScriptShiftLayeredInterface2025b}.
Recent work on just-in-time objectives further shows how AI systems can infer a user's in-the-moment goals from observed context, such as screenshots and webpage text, and use these objectives to generate specialized expert responses or customized tools for the user's current task~\citep{lamJustInTimeObjectivesGeneral2025}.
These systems show that customization can help writers access diverse perspectives, delegate specialized tasks, generate situated tools, and better align AI support with their communicative intentions.

However, existing customizable writing agents are primarily invoked on demand: writers can define who the AI should be, what it should do, or what objectives it should optimize for, but they must still invoke it manually. Our work introduces a new customization dimension: the \emph{proactivity} of AI writing agents. In our probe, writers can shape both the type of cognitive support they receive and the moments when that support enters the writing process. These user configurations provide unique empirical insights into how writers imagine, configure, and bound proactive AI support, helping us map the broader design space of proactive cognitive support for writing.\looseness=-1

%% file: sections/4-system-design.tex
\section{Design Goals}
We designed our system as a technology probe: \textit{an instrument deployed to investigate an open-ended design space and return useful or interesting data about users' practices, needs, and values}~\citep{hutchinsonTechnologyProbesInspiring2003}.
Thus, rather than aiming to build an optimal writing system, our goal was to create a functional, deployable writing environment in which users could create and use proactive thought partners for their own writing tasks.
This allowed us to surface insights into how users would appropriate proactive thought partners and how they would experience customizable, proactive AI support during writing.
Specifically, we designed the probe around three goals.

\namedparagraph{[DG1] Support user-configurable proactive partners}
We use customization as a lens to understand how users expect proactive thought partners to support their writing practices and needs.
Based on prior work on customizable AI agents~\citep{zhengCustomizingEmotionalSupport2025,lehmannCollaborativeDocumentEditing2026,benharrakWriterDefinedAIPersonas2024a} and mixed-initiative systems~\citep{horvitzPrinciplesMixedinitiativeUser1999a,puAssistanceDisruptionExploring2025a,baileyEffectsInterruptionsTask2001,baileyMeasuringEffectsInterruptions2000,chenNeedHelpDesigning2025a,luoHowUsersPerceive2026}, we focus on two basic forms of user customization: the role and proactivity of partners.
The role specifies what kind of support the partner should provide, a dimension emphasized in much prior work on customizable AI agents.
The proactivity, by contrast, specifies when and under what writing contexts the partner should intervene, including the moments that should be treated as candidates for support and the contextual conditions under which the partner should take the initiative.

\namedparagraph{[DG2] Enable flexible engagement with proactive support}
Proactive support can arrive at moments when users may or may not want assistance.
At different moments, users may also desire different levels of AI involvement in their writing process (\eg inspiration or direct text generation).
Thus, the probe needs to preserve user agency over how AI engages in writing~\citep{horvitzPrinciplesMixedinitiativeUser1999a}.
Informed by prior work on human-AI co-writing~\citep{carreraWhereAddEgg2025,fundalDirectionalAlignmentNarrative2026,yehGhostWriterAugmentingCollaborative2025,yinProactiveAICatalyst2026,draxlerAIGhostwriterEffect2024}, we investigate three forms of engagement: users can ignore a suggestion, use it as inspiration, or ask the partner to apply changes directly to the document.
Together, these options allow users to treat proactive suggestions as peripheral cues, sources of inspiration, or actionable edits, while allowing us to observe how users dismiss, accept, or use proactive AI support in practice.\looseness=-1

\namedparagraph{[DG3] Embed proactive support in a lightweight writing environment}
To study proactive thought partners in use, the probe needs to support writing as an ongoing activity rather than a one-off interaction with an assistant.
We therefore aim to embed proactive support in a lightweight but functional and deployable writing environment that provides the basic affordances needed for composing and revising text while minimizing additional interface complexity.
This allows users to engage with proactive support during their own writing tasks without making the writing environment itself the primary focus of the study.
At the same time, the environment should enable the partners to monitor writing traces, detect candidate moments for intervention, and surface proactive support in context.\looseness=-1


\section{Proactive Thought Partners: A Technology Probe}
In this section, we describe the design and implementation of our technology probe, guided by the design goals above and inspired by Horvitz's principles of mixed-initiative user interfaces~\citep{horvitzPrinciplesMixedinitiativeUser1999a}.
The probe consists of three main interfaces.
First, an onboarding panel allows users to log in and describe their writing goals for a writing session.
Second, a partner configuration panel allows users to create, configure, and enable proactive thought partners (\figref{fig:config}).
Third, a lightweight Markdown editor allows users to write with their enabled partners (\figref{fig:editor}).
The following subsections describe the probe in four parts: how users create and configure proactive thought partners (\secref{sec:partner_custom}), how the probe activates relevant partners at opportune moments in the ongoing writing process (\secref{sec:partner_timing}), how the activated partners generate thought-provoking suggestions (\secref{sec:partner_suggestion}), and how users can engage with proactive suggestions once they appear (\secref{sec:partner_engagement}).\looseness=-1

\input{figures/fig-editor}
\input{figures/fig-config}

\subsection{Partner Customization}
\label{sec:partner_custom}
In the configuration panel, users can create a proactive thought partner by specifying a name and emoji, a role, one or more event triggers, and a contextual heuristic.
Together, these fields allow users to customize both the partner's role and its proactivity (\textbf{DG1}): what kind of support the partner should provide, when it may intervene, and under what contextual conditions its intervention may be useful.\looseness=-1

\begin{packeditemize}
    \item \textbf{Partner Role} defines the partner's expected responsibility and capability, such as helping generate ideas, identifying weak evidence, or challenging an argument. This field follows prior work on customizable AI agents by allowing users to define the kind of support they want from an AI collaborator~\citep{zhengCustomizingEmotionalSupport2025,lehmannCollaborativeDocumentEditing2026,benharrakWriterDefinedAIPersonas2024a}.
    \item \textbf{Event Trigger(s)} define an observable user action that may indicate an opportunity for proactive support, such as pausing, completing a sentence, or selecting text. Inspired by prior work on proactive AI~\citep{puAssistanceDisruptionExploring2025a,baileyEffectsInterruptionsTask2001,baileyMeasuringEffectsInterruptions2000,chenNeedHelpDesigning2025a,luoHowUsersPerceive2026}, this field allows users to specify broad candidate moments when a partner may consider intervening. Users can select one or more triggers for each partner.
    \item \textbf{Contextual Heuristic} defines the contextual criteria under which the partner should take the initiative following a triggering event. Drawing on work in context-aware computing~\citep{deyUnderstandingUsingContext2001,byunHarnessingContextSupport2002}, this field reflects how users expect the proactive system to interpret the writing context and make judgments. By specifying contextual heuristics, users define \textit{specific} timing conditions (such as when a claim lacks evidence or an argument may need a counterpoint) under which a particular partner would be relevant and helpful when \textit{broad} event triggers occur.
\end{packeditemize}

\subsection{Partner Activation}
\label{sec:partner_timing}

Event triggers and contextual heuristics together determine when a partner may step in (\figref{fig:pipeline}): triggers identify broad candidate moments for intervention, while heuristics determine whether a partner is relevant to the current writing context in a more precise way.
First, we implemented three event triggers, each grounded in a different rationale from prior work on proactive AI~\citep{puAssistanceDisruptionExploring2025a,yinProactiveAICatalyst2026,chenNeedHelpDesigning2025a}, interruption management~\citep{baileyEffectsInterruptionsTask2001,czerwinskiInstantMessagingEffects2000,iqbalIndexOpportunity2005,normanControlMultipleActivities1986}, and keystroke analysis of writing~\citep{baaijenKeystrokeAnalysisReflections2012,matsuhashiPausingPlanningTempo1981,schilperoordItsTimeTemporal2022,galbraithAligningKeystrokesCognitive2019}.
These triggers are rule-based and operate on user keystrokes monitored in real time within the editor.
\looseness=-1

\begin{packeditemize}
    \item \namedparagraph{Long Pause} The first rationale is that opportune moments for interruption often occur during natural breaks in user activity~\citep{baileyEffectsInterruptionsTask2001,baileyMeasuringEffectsInterruptions2000,czerwinskiInstantMessagingEffects2000}. In writing, pauses have long been seen as natural breaks in writers' cognitive processes~\citep{chenuInterwordIntrawordPause2014,barkaouiWhatCanL22019,medimorecPausesWrittenComposition2017}. In our probe, the pause trigger fires after a period of user inactivity in the editor, set to 5 seconds by default and customizable by users.
    \item \namedparagraph{Sentence End} The second rationale is that interventions are often perceived as less disruptive at task or subtask boundaries~\citep{czerwinskiInstantMessagingEffects2000,iqbalIndexOpportunity2005,normanControlMultipleActivities1986}. We operationalize this rationale through a sentence-end trigger, because completing a sentence creates a natural boundary in the flow of writing. 
    To avoid interrupting rapid typing, the system waits for a short idle period after sentence completion, set to 1 second by default and customizable.\looseness=-1
    \item \namedparagraph{Text Selection} The third rationale draws from proactive AI tools that treat users' implicit actions as signals of possible need~\citep{puAssistanceDisruptionExploring2025a,namUsingLLMHelp2024b}. 
    Although text selection is a more explicit signal than passive inactivity, we followed recent work on proactive AI~\citep{puAssistanceDisruptionExploring2025a} and included it as a trigger because it still enables a lightweight form of proactivity: the writer does not need to formulate a prompt or select a specific command for a partner to offer support.
    When a user selects text and remains idle for a short period, set to 5 seconds by default and customizable by users, the system treats the selected span as the local focus for possible partner intervention.
\end{packeditemize}

\input{figures/fig-pipeline}

When an event trigger occurs, the system sends the writer's session goal, the current text in the editor, the recent writing behaviors, and the list of enabled partners associated with that event trigger to an LLM-based decision engine (\figref{fig:pipeline}).
The writing behaviors include the current cursor position, the trigger event, and the keystroke logs from the 15 seconds\footnote{We use a 15-second window, following prior keystroke-analysis studies that used similarly short windows to infer user states such as boredom and engagement~\citep{kuvarAutomaticallyDetectingTaskunrelated2023,bixlerDetectingBoredomEngagement2013}.} before the trigger.
Based on this contextual information, the decision engine selects at most two enabled partners, if any, to intervene at that moment by checking whether their user-defined contextual heuristics are satisfied.
Activated partners appear as small floating tags in the right-side panel (\figref{fig:engagement}A.1), aligned with the user's current cursor position.
Each tag displays the partner's emoji and name, providing a lightweight peripheral cue before users decide how to engage with it (\secref{sec:partner_engagement}).\looseness=-1

\subsection{Suggestion Generation}
\label{sec:partner_suggestion}

The activated partner will generate a suggestion based on the current writing context.
Users see this suggestion after they click the floating tag (\figref{fig:engagement}C).
Each suggestion contains two components.

\begin{packeditemize}
    \item \textbf{Acknowledgement}: First, the partner briefly acknowledges what the writer has just done, is currently doing, or may be trying to do next. This component is inspired by writing-feedback research emphasizing that effective feedback should first locate itself in the writer's current draft and make visible what the responder understands the writer to be attempting, before offering guidance for next steps~\citep{sommersRespondingStudentWriting1982a,nicolFormativeAssessmentSelfregulated2006a}.
    \item \textbf{Suggestion}: Second, the partner offers a thought-provoking suggestion tailored to the current context. We frame this suggestion as a question because prior work suggests that question-style AI suggestions can stimulate high-quality ideas while preserving users' ownership~\citep{maierPartneringGenerativeAI2026}.\looseness=-1
\end{packeditemize}

For example, the Evidence Partner responds to a writer who has just introduced the importance of elite players' training habits:

\begin{quote}
\textit{It looks like you are moving from general principles to concrete examples from elite players.
How might you introduce a specific anecdote or training philosophy from a top player, such as Rafael Nadal's relentless drills or Roger Federer's recovery routines, to illustrate the principles you just mentioned?}
\end{quote}

\input{figures/fig-engagement}

\subsection{Engagement Forms}
\label{sec:partner_engagement}

Once the floating tags of activated partners appear, users can choose whether and how to engage with the proactive interventions in three ways (\textbf{DG2}).

\begin{packeditemize}
    \item {\textbf{Ignoring}} (\figref{fig:editor}B):
    If writers do not need support at that moment, they can simply continue writing.
    The floating tag gradually fades out and disappears after 15 seconds.
    This allows proactive support to remain optional and low-commitment: the system can make a bid for the writer's attention without requiring an explicit dismissal or interrupting the writing flow.
    This also reflects the principle of allowing efficient direct termination for mixed-initiative user interfaces~\citep{horvitzPrinciplesMixedinitiativeUser1999a}.\looseness=-1
    \item{\textbf{Inspiring}} (\figref{fig:editor}C):
    Users can click the floating tag to reveal a suggestion.
    The tag expands into a floating card at the same position displaying the partner's suggestion.
    The suggestion serves as inspiration for the user to continue developing or refining their prose.
    Users can also click a message icon to engage in a follow-up conversation with the partner.
    This allows them to clarify the suggestion, ask for alternatives, challenge the partner's framing, or explore related ideas before continuing.
    \item{\textbf{Executing}} (\figref{fig:editor}D):
    If writers decide the suggestion should be incorporated into the document, they can optionally ask the partner to execute it by inserting or revising text directly in the editor.
    Executing represents the deepest form of engagement in our probe: the partner moves from offering cognitive scaffolding to temporarily taking over part of the writing work, while the writer retains control by explicitly initiating the action and reviewing the resulting change before accepting it.
\end{packeditemize}

Together, these three forms of engagement represent increasing levels of AI involvement.
Writers can ignore a partner's intervention, treat it as inspiration, or delegate concrete writing actions.
This graduated interaction design aims to preserve writer agency while still allowing proactive partners to provide more substantive support when writers want it.
At the same time, these engagement options serve a probing function: by observing whether writers ignore, open, discuss, or execute proactive suggestions, we can examine how they prefer AI to engage in their writing process.\looseness=-1

\subsection{Implementation}
\label{sec:implementation}
Our probe is implemented with Next.js\footnote{\url{https://nextjs.org/}}, which supports server-side rendering and backend API routes for calling external services, including the Google Gemini APIs\footnote{\url{https://ai.google.dev/gemini-api/docs}} for LLM-based functions and Firebase APIs\footnote{\url{https://firebase.google.com/}} for logging user events.
The Markdown editor is built with BlockNote\footnote{\url{https://www.blocknotejs.org/}}.
We also adapted BlockNote's AI SDK integration to support partners' text insertion and revision in the editor when invoked by users.
Keystroke logging is implemented with JavaScript event listeners, following prior implementation practices from~\citet{tianKLiCKeCorpusKeystroke2025}.
For the decision engine that selects which partners should intervene, we use \texttt{gemini-2.5-flash-lite} to minimize latency. In practice, the decision engine takes less than one second to select partners after an event is triggered. The partner's floating tag appears immediately (\figref{fig:engagement}) after the decision is made to reduce participants' perceived latency, while the partner agent takes approximately eight seconds to generate a suggestion.
For other LLM-based tasks, including suggestion generation, follow-up discussion, and text insertion or revision, we use \texttt{gemini-2.5-flash}.
The prompts are provided in the supplementary materials.

%% file: figures/fig-editor.tex
\begin{figure*}
\centering
\includegraphics[width=\linewidth]{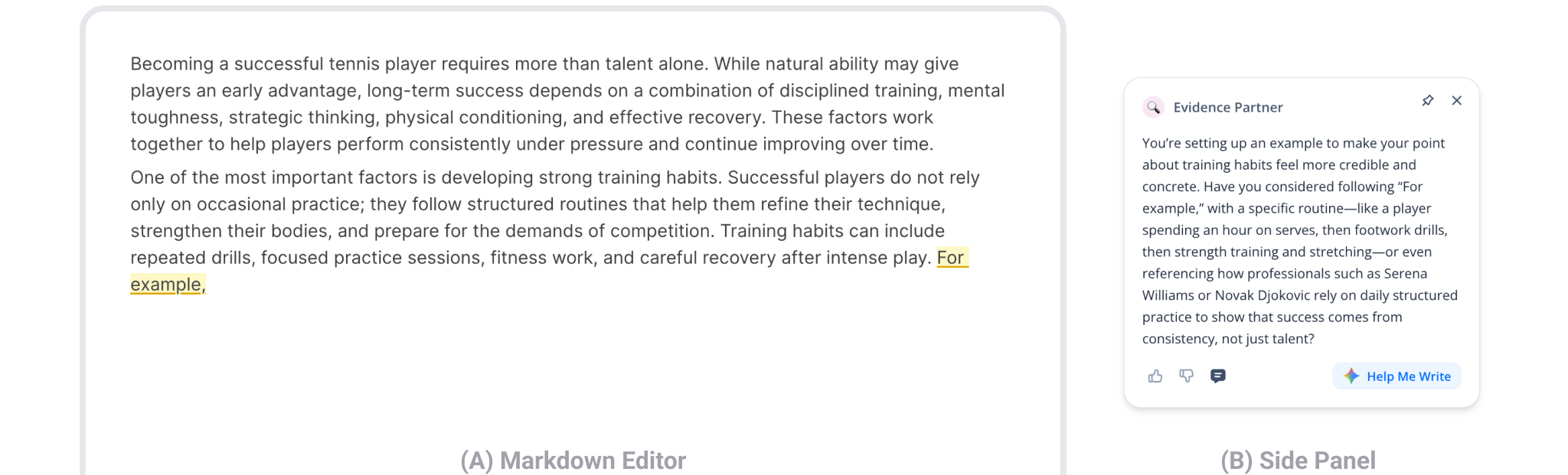}
\caption{\formatcaption{Writing environment with proactive thought partners.}{The probe consists of a lightweight Markdown editor (A) and a side panel (B) for proactive suggestions. As writers compose text, relevant partners take the initiative to surface contextually relevant suggestions alongside the draft. Here, the Evidence Partner notices an unfinished example and suggests concrete ways to support the claim about training habits.}}
\label{fig:editor}
\end{figure*}

%% file: figures/fig-config.tex
\begin{figure}
\centering
\includegraphics[width=\linewidth]{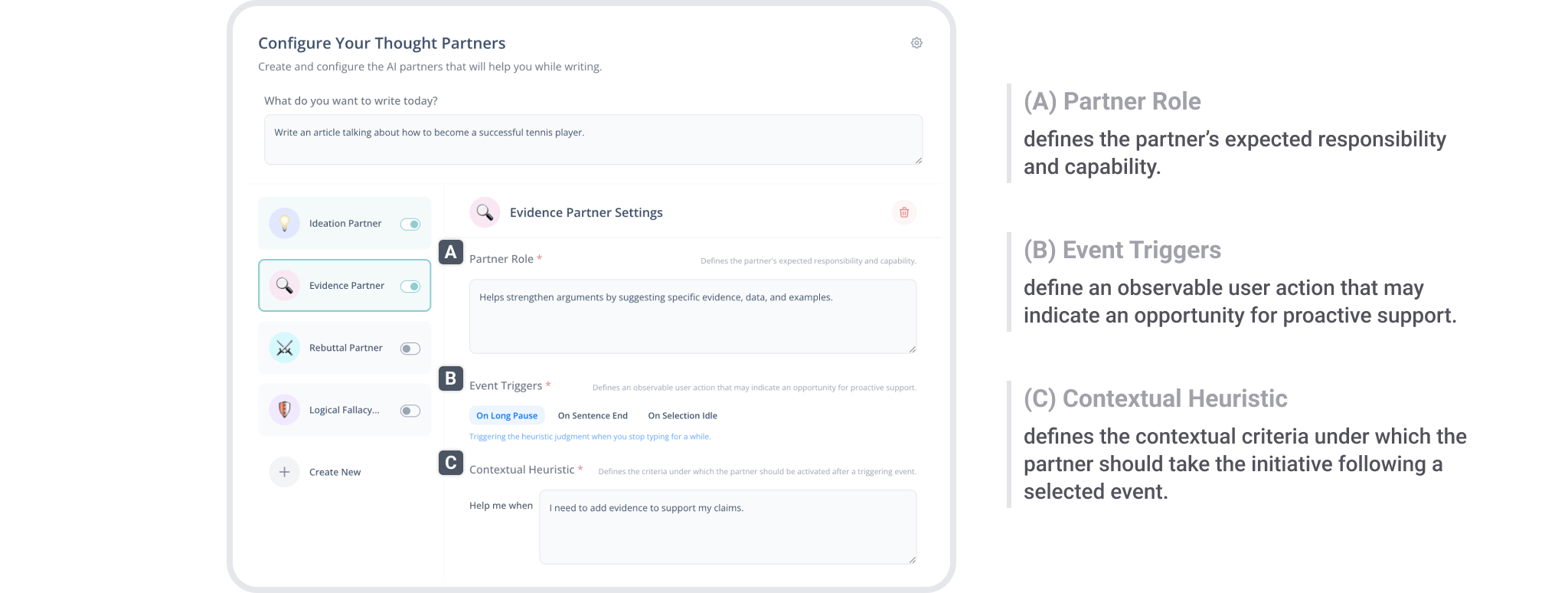}
  \caption{\formatcaption{Configuring proactive thought partners.}{The configuration panel allows users to design and enable their own proactive thought partners. Users first specify a name and emoji for a partner, then define the role (A), select event triggers (none is selected by default to avoid anchoring effects), (B), and specify a contextual heuristic that determines when the partner should take the initiative after a selected event (C). In this example, the Evidence Partner is configured to help strengthen claims by suggesting evidence when the writer pauses and the current context calls for examples.}}
\label{fig:config}
\end{figure}



%% file: figures/fig-pipeline.tex
\begin{figure*}
\centering
\includegraphics[width=\linewidth]{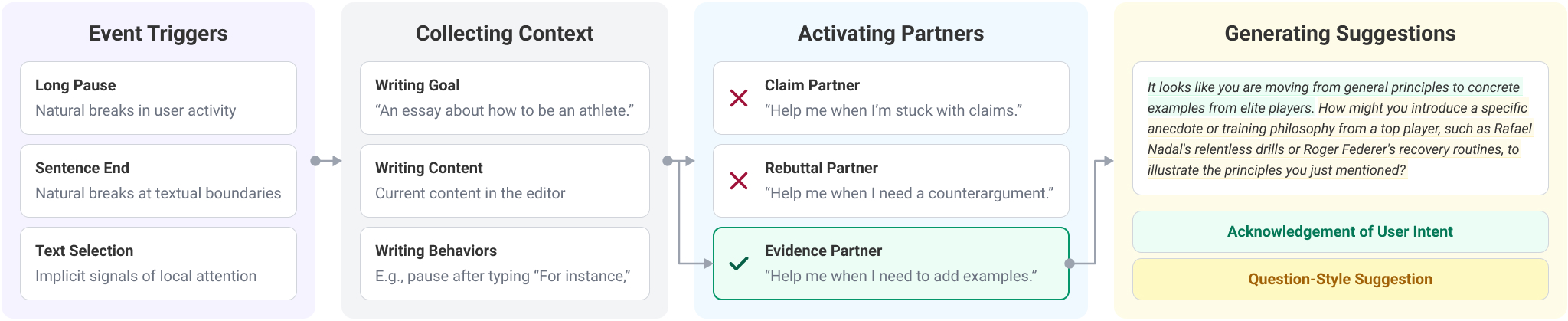}
\caption{\formatcaption{System pipeline for proactive thought partners.}{When an event trigger (\eg a pause, sentence end, or text selection) occurs, the system gathers contextual information about the writer's session goal, current text, and writing behaviors, including keystroke logs. It then evaluates the user-defined contextual criteria of all enabled partners and activates the partner or partners, if any, whose criteria best match the current situation. Each activated partner generates a suggestion that includes both an acknowledgment of the inferred user intent and a question-style suggestion based on its role.}}
\label{fig:pipeline}
\end{figure*}

%% file: figures/fig-engagement.tex
\begin{figure*}
\centering
\includegraphics[width=\linewidth]{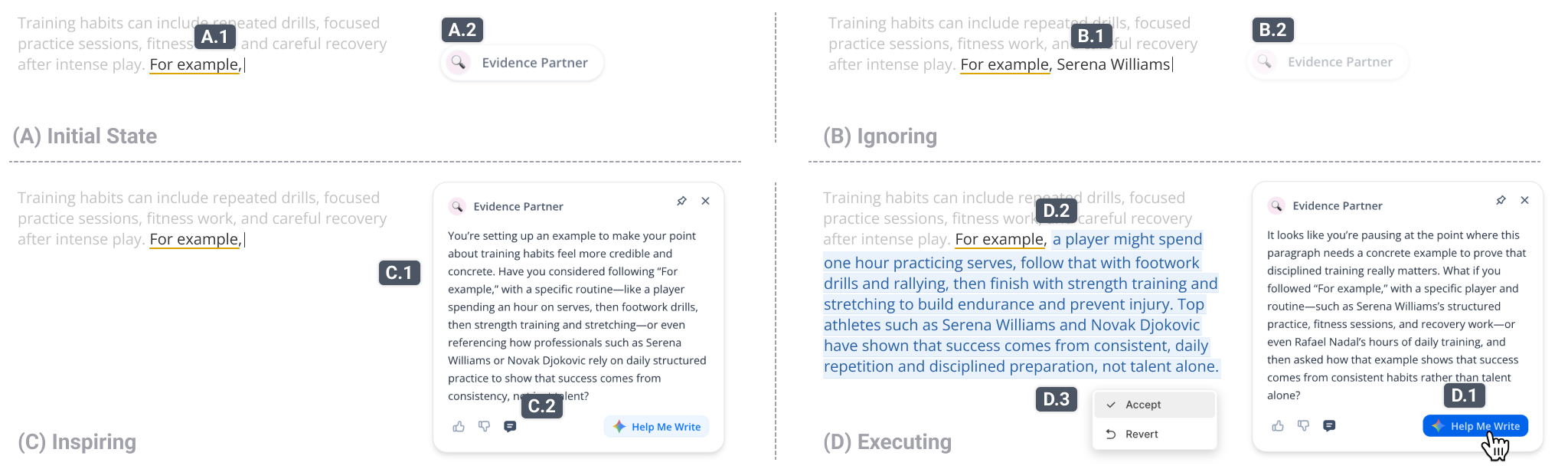}
\caption{\formatcaption{Forms of engagement with proactive suggestions.}{When a writer pauses after typing (A.1), an activated partner first appears as a lightweight floating tag (A.2). If the writer ignores the tag and continues writing (B.1), it gradually fades out and disappears (B.2). If the writer clicks the tag, it expands into a suggestion card (C.1), where the writer can inspect the suggestion and optionally click the message icon (C.2) to ask follow-up questions. If the writer hovers over the suggestion card and clicks the \textit{Help Me Write} button (D.1), the partner directly inserts or revises text in the editor (D.2), after which the writer can accept or revert the change (D.3).}}
\label{fig:engagement}
\end{figure*}

%% file: sections/5-evaluation.tex
\section{User Study}
We conducted a one-week technology probe study with 16 participants to explore how writers customize, experience, and engage with proactive thought partners in writing practice.
It is important to note that a technology probe study is not an evaluative study of a finished product's efficacy.
Instead, it is an exploratory deployment intended to inspire new design directions and uncover how users adapt to, appropriate, and make sense of a nascent technology in practice~\citep{kaur_interpreting_2020,seymour_informing_2020,luAwarenessActionExploring2024}.
In this spirit, our deployment aimed to investigate the following research questions:\looseness=-1

\begin{itemize}
    \item[\textbf{RQ1}] \textit{Partner Creation.} How do users create and configure their proactive thought partners to suit their writing needs?
    \item[\textbf{RQ2}] \textit{Suggestion Engagement.} How do users engage with proactive suggestions from these partners during writing?
    \item[\textbf{RQ3}] \textit{User Experience.} How do users experience this new form of customizable, proactive cognitive support for writing?\looseness=-1
\end{itemize}

\input{tables/tab-participants}

\subsection{Participants}
We recruited 16 participants through an organizational mailing list with more than 10,000 recipients at a large global software company (Table~\ref{tab:participants}). Participants ranged in age from 26 to 53 years ($M = 37.47$, $SD = 8.57$; one participant preferred not to disclose their age), with 11 identifying as female and 5 as male. Because the study involved open-ended writing tasks in English, we required participants to be proficient in reading and writing in English; 5 were native speakers and 11 were non-native speakers. We did not stratify recruitment by level of professional writing expertise. Instead, we sought everyday writers with regular writing practices rather than focusing exclusively on expert or novice writers. Participants therefore came from diverse professional backgrounds, including content strategy, communications, software engineering, and program management.

To ensure that participants could meaningfully engage with and reflect on the probe, we purposively sampled individuals with prior experience using AI tools for writing. During the three months preceding the study, 10 participants reported using AI tools on a daily basis, and 6 used them weekly. Participants shared their detailed experiences with AI-assisted writing, including specific tools used and their primary writing purposes. In summary, participants used various AI tools such as ChatGPT, Gemini, Claude, and Smart Compose in Gmail. They applied these tools to a wide range of tasks, including drafting technical and professional documentation (\eg daily reports, policy documents, research papers), managing routine correspondence, and engaging in personal or creative writing (\eg fiction, journaling, scripts). Across these contexts, participants used AI throughout the writing process to brainstorm outlines, draft content, and iteratively edit for tone, clarity, and conciseness.


\subsection{Procedure}
\label{sec:study_procedure}
The study consisted of three phases: an onboarding session, a week-long diary study, and a concluding semi-structured interview.
Before the onboarding session, participants provided informed consent and completed a demographics questionnaire. 
After the study, each participant was compensated \$140 for their participation.

\namedparagraph{Phase 1: Onboarding (60 minutes)}
We conducted a remote onboarding session with each participant.
Participants first received a short tutorial on the probe's core features.
To help participants become familiar with the concept of proactive thought partners, we also provided four preconfigured partners during onboarding: an \textit{Ideation Partner}, an \textit{Evidence Partner}, a \textit{Rebuttal Partner}, and a \textit{Logical Fallacy Partner}.
These partners were designed around the warm-up writing task below and informed by prior work on AI-assisted argumentative writing, such as VISAR~\citep{zhangVISARHumanAIArgumentative2023}.
Details of these preconfigured partners, including their configurations, are provided in the supplementary materials.\looseness=-1

After tutorial, they completed a warm-up writing task, in which they wrote a short essay of approximately 300 words on the positive and negative aspects of AI in daily life.
This task was designed to familiarize participants with the probe before the diary study.
During the warm-up task, participants were asked to (1) create two custom thought partners and optionally enable or update any preconfigured partners, and (2) complete the essay with their enabled partners to observe how proactive suggestions were triggered and to practice key interaction features.
After the warm-up session, we provided instructions for the diary study.\looseness=-1

\namedparagraph{Phase 2: Diary Study (7 days)}
Participants began using the probe the following day for seven days.
To capture how participants appropriate proactive thought partners in their own writing practices, we did not constrain the genre, topic, or purpose of their writing.
However, we asked participants to complete at least four substantive writing sessions during the week to ensure sufficient exposure to the probe.
We defined a \textit{substantive session} as one that met two criteria: (1) at least 20 minutes of active writing or production of at least 300 words; and (2) completion of a self-contained piece of writing, rather than randomly connected sentences or artificial text written only for the study.
To mitigate privacy risks associated with system logging, we explicitly instructed participants not to enter personally identifiable information, confidential workplace information, or other highly sensitive data into the probe.\looseness=-1

During deployment, the probe collected interaction logs, including partner configurations, keystroke logs, and user interactions with proactive suggestions.
For each proactive intervention they received, participants could optionally provide binary feedback by clicking a thumbs-up or thumbs-down icon (\figref{fig:editor}B).
In addition, participants submitted a structured diary entry at the end of each writing session.
The diary included three 10-point scales measuring satisfaction, perceived timeliness, and perceived helpfulness of the proactive support, as well as four open-ended prompts asking about (1) the partners they configured and why, (2) the proactive support they received and their feedback on it, (3) challenges encountered while using the probe, and (4) any other observations.
Together, these data sources allowed us to triangulate automated system logs, immediate feedback on individual  interventions, and longer-form reflections on each writing session.\looseness=-1

\namedparagraph{Phase 3: Semi-Structured Interview (60 minutes)}
In the final phase, participants first completed the short-form Usability Metric for User Experience (UMUX-LITE)~\citep{lewisUMUXLITEWhenTheres2013}.
We then conducted a semi-structured interview to probe participants' week-long experiences and elicit their visions for future proactive thought partners for writing.
We began by asking participants to describe their general experience using the probe during the week.
We then asked them to compare the proactive AI features in our probe with their experiences using reactive AI chatbots, such as ChatGPT or Gemini, and proactive next-phrase suggestions, such as Smart Compose in Gmail~\citep{chen_gmail_2019}.
Next, we used diary entries, binary feedback logs, and system logs as elicitation artifacts, asking participants to elaborate on their strategies for configuring thought partners, specific instances in which proactive suggestions were helpful or intrusive, and their rationales for how they chose to engage with the proactive support they received.
We also explored perceived benefits and challenges of working with proactive thought partners, as well as participants' perceived ownership and agency during the writing process.
Building on these reflections, we shifted the discussion beyond the current prototype's constraints and asked participants to articulate their ideal future proactive writing assistants.
The interview protocol is provided in the supplementary materials.
\looseness=-1

\subsection{Analysis}
We analyzed multiple sources of qualitative and quantitative data, including partner configurations, interaction logs, diary entries, and semi-structured interview transcripts. The interaction logs were used to characterize participants' partner creation behaviors and engagement with proactive interventions, while diary entries and interviews provided insight into participants' rationales, experiences, and perceptions. Two researchers conducted a thematic analysis of the qualitative data~\citep{braunSuccessfulQualitativeResearch2013}.
They reviewed the interview transcripts and diary entries to identify recurring concepts, then met to compare, discuss, and consolidate the codes into a codebook. 
The researchers then collaboratively grouped related codes into higher-level themes that addressed our research questions through multiple rounds of discussion. 
Throughout the analysis, interaction logs and quantitative usage metrics were used to triangulate and contextualize the qualitative findings.\looseness=-1

%% file: tables/tab-participants.tex
\begin{table*}
\footnotesize
\centering
\caption{\formatcaption{Participant demographics and writing activity.}{This table presents participants' ages, genders, role, whether they are native English speakers (NES), self-reported frequency of AI usage for writing prior to the study, number of completed writing sessions, and writing contexts across sessions. Writing contexts are categorized based on the taxonomy (\ie Academic, Creative, Journalistic, Technical, Professional, Personal) proposed by~\citet{leeDesignSpaceIntelligent2024a}.}}
\label{tab:participants}
\begin{tblr}{
  colspec = {lllllll},
  hline{1,Z} = {0.8pt, solid},
  hline{2} = {0.4pt, solid},
  row{1} = {bg=gray!10},
  cell{2-17}{1} = {r=1}{font=\bfseries,bg=gray!6},
}
\textbf{ID} & \textbf{Gender} & \textbf{Age} & \textbf{Job Title}  & \textbf{NES?} & \textbf{AI Usage} & \textbf{Writing Sessions and Contexts}\\
\textbf{P01} & Male & 45 & Software Engineer & No & Daily & Creative, Professional, Personal, Technical \\
\textbf{P02} & Male & 28 & Software Engineer & No & Weekly & Academic (3), Creative \\
\textbf{P03} & Female & --- & Program Manager & No & Daily & Academic (2), Creative, Personal \\
\textbf{P04} & Female & 44 & Program Manager & No & Daily & Academic (2), Creative, Professional \\
\textbf{P05} & Female & 29 & Software Engineer & No & Weekly & Academic (2), Personal, Creative \\
\textbf{P06} & Male & 37 & Customer Engineer & Yes & Daily & Technical (2), Professional (2) \\
\textbf{P07} & Female & 34 & Software Engineer & No & Weekly & Personal (3), Technical \\
\textbf{P08} & Female & 31 & Commercial Strategist & Yes & Weekly & Personal (3), Academic \\
\textbf{P09} & Female & 38 & Content Strategist & Yes & Daily & Academic (2) Journalistic, Creative \\
\textbf{P10} & Female & 29 & Software Engineer & Yes & Daily & Academic (4) \\
\textbf{P11} & Female & 26 & Systems Engineer & No & Weekly & Personal (2), Professional, Journalistic \\
\textbf{P12} & Male & 38 & Pricing Specialist & No & Weekly & Professional (2), Creative, Journalistic \\
\textbf{P13} & Female & 42 & Program Manager & Yes & Daily & Personal (3) Creative \\
\textbf{P14} & Female & 35 & Communications Manager & No & Daily & Personal (2) Academic, Technical \\
\textbf{P15} & Female & 53 & Executive Operations Lead & No & Daily & Personal (2), Academic (2) \\
\textbf{P16} & Male & 53 & Application Engineer & No & Daily & Technical (3), Professional (2), Personal \\
\end{tblr}
\end{table*}

%% file: sections/6-results.tex
\section{Findings}
In this section, we present findings related to our research questions. We first provide an overview of participants' use of the probe during the one-week deployment and their perceptions of its usability. We then examine how participants created proactive thought partners to anticipate writing needs (RQ1), how they engaged with proactive suggestions during writing (RQ2), and how they perceived the customized, proactive cognitive support (RQ3).

\subsection{Probe Usage and Usability}

Across the 16 participants, we collected 66 writing sessions, excluding onboarding sessions.
Participants completed an average of 4.13 ($SD=0.50$) sessions each.
Each session produced an average of 390.34 ($SD=151.41$) words and lasted an average of 17.09 ($SD=7.47$) minutes.
Participants used the probe for a wide range of writing contexts.
Based on the taxonomy from~\citet{leeDesignSpaceIntelligent2024a}, these sessions included personal writing ($N=19$), academic writing ($N=19$), professional writing ($N=9$), creative writing ($N=8$), technical writing ($N=8$), and journalistic writing ($N=3$).
The writing sessions completed by each participant are summarized in \tabref{tab:participants}.

The System Usability Scale (SUS) scores calculated from UMUX-LITE collected during the exit interviews averaged 82.81 ($SD=11.97$).
This score corresponds to the ``Excellent'' category defined by~\citet{bangorDeterminingWhatIndividual2009,bangorEmpiricalEvaluationSystem2008}, placing it in approximately the 90th-95th percentile.
This is consistent with daily survey measures of overall satisfaction, which averaged 8.16/10 ($SD=1.76$).
A linear mixed-effects model with participant as a random effect showed that satisfaction increased significantly with session index ($b = 0.33$, $SE = 0.16$, $p = .036$), indicating that participants became progressively more satisfied with the probe as they gained experience over the course of the deployment.
\looseness=-1

\subsection{RQ1: Partner Creation}

Participants created 54 partners, with an average of 3.38 ($SD=1.96$) partners created per participant.
Among these, 35 (64.81\%) were enabled in only one writing session, while 19 (35.19\%) were used across multiple sessions.
Table~\ref{tab:representative_partners} shows examples of these partners.
We next examine the overarching strategies participants used to create partners (\textbf{KF1}), the types of roles these partners took on (\textbf{KF2}), and how participants configured their partners' proactivity (\textbf{KF3}).

\namedparagraph{KF1: Creating partners by anticipating writing needs}
\label{sec:rq1_strategy}
Participants configured partners prospectively, reasoning about what cognitive support they would need before writing began. 
This took two forms. In \textbf{goal-driven design}, participants started from a mental image of the finished piece and worked backward to infer the expertise it required. P04 described the approach directly: ``\textit{I try to envision what the end product would look like and then, if I had a human writer, I imagine what that person's expertise should be.}'' 
In \textbf{difficulty-driven design}, participants instead began from self-knowledge of their own recurring writing difficulties and designed partners to preemptively address those gaps. 
P13 articulated this explicitly: ``\textit{I knew myself well enough to know what potential blockers I would have towards reaching that goal\dots~where I anticipated myself struggling.}'' 
Both strategies required participants to mentally simulate the writing task before it began: either from the perspective of the intended output or from the perspective of their own likely difficulties.\looseness=-1

\input{tables/tab-partners}

\namedparagraph{KF2: Partner roles clustered around distinct forms of cognitive support}
Participants configured 43 of 54 partners (79.63\%) for higher-level cognitive support and 11 (20.37\%) for local \textbf{textual assistance} such as language and style refinement. This pattern aligns with our expectation that proactive thought partners would primarily support cognitive activities beyond phrasing and text production.
Specifically, 14 partners (25.93\%) focused on seeking evidence or domain knowledge (\textbf{information seeking}), 12 (22.22\%) on argument development and reasoning (\textbf{argument development}), 10 (18.52\%) on perspective-taking and critical reflection (\textbf{critical reflection}), and 7 (12.96\%) on ideation and creative development (\textbf{ideation support}).
These roles show that participants envisioned proactive partners as specialized sources of knowledge, reasoning support, reflective scaffolds, alternative perspectives, and inspiration that could assist different forms of cognitive work during writing.\looseness=-1

\namedparagraph{KF3: Partner proactivity is configured through broad event triggers and context-specific heuristics.}
We then analyzed how participants configured event triggers and contextual heuristics to control partners' proactivity. We found that event triggers were generally configured broadly: 31 of 54 partners (57.41\%) were enabled with all three triggers. 
Specifically, \textbf{pauses} were enabled for 40 of 54 partners (74.07\%), \textbf{sentence endings} for 40 partners (74.07\%), and \textbf{text selections} for 44 partners (81.48\%).
In contrast, participants used contextual heuristics to specify more particular writing situations in which a partner's support would be relevant. We identified three recurring forms of contextual cues. First, \textbf{draft-state cues} described what was currently present in the text, such as language being too casual, an argument presenting only one perspective, or the writing becoming repetitive or boring 35/54 (64.81\%). Second, \textbf{writing-activity cues} described what the writer was currently doing in the writing process, such as introducing a concept, developing a claim, analyzing an issue, or moving from listing arguments toward a conclusion 38/54 (70.37\%). Third, \textbf{anticipated-need cues} described the support the writer might need at that moment or next, such as an example, another perspective, clarification of a concept, or a stronger argument 36/54 (66.67\%).
Together, these patterns suggest a division of labor between the two timing mechanisms: event triggers specified \textit{when the system should look}, while contextual heuristics specified \textit{what it should look for} when determining whether and which partner should intervene. 
Participants viewed interaction events as broader opportunities for support, while whether support was actually needed depended on how those events were interpreted in relation to the writer's evolving state and goals.


\begin{takeawaybox}{RQ1 Takeaway}
\textit{Participants configured proactive thought partners through a form of prospective planning. They anticipated the goals and difficulties of an upcoming writing task, translated these expectations into specialized partner roles, and specified when those roles might become relevant. While event triggers were configured broadly as candidate moments for intervention, contextual heuristics encoded more specific expectations about the evolving draft, current writing activity, and anticipated support needs.}\looseness=-1
\end{takeawaybox}


\subsection{RQ2: Suggestion Engagement}
The probe provided 1,100 proactive interventions across all participants and their writing sessions (\tabref{tab:interaction_metrics}).
Among these, participants opened 41.73\% of the surfaced interventions (459/1,100).
Below, we first present our findings on how participants leveraged proactive suggestions in their writing process (KF4). We then examine why they ignored many surfaced interventions (KF5) and why, in some cases, they asked the AI to take over and execute edits directly in the document (KF6).\looseness=-1

\input{tables/tab-engagement}

\namedparagraph{KF4: Suggestions supported both idea generation and self-monitoring}
Participants described using proactive suggestions primarily in two ways. First, suggestions were generative: they introduced ideas, connections, or directions that writers had not reached on their own. The most memorable cases involved unexpected connections that extended participants' thinking. P12 described a suggestion linking Buddhist and Abrahamic religious concepts as ``\textit{an eye opener I hadn't thought about it},'' while P01 recalled a suggestion connecting database behavior to memory leak debugging: ``\textit{I never related the database to memory leak\dots it just gave me some suggestion and then you know it became like a kind of my idea right.}''\looseness=-1

Second, suggestions were regulatory: they helped writers assess whether their writing remained aligned with their goals and arguments. P13 valued prompts that surfaced when they were ``\textit{veering off track},'' and prompting them to reflect on whether their goal had changed.
Others valued suggestions that surfaced gaps or neglected perspectives.
For P06, even a simple prompt such as ``\textit{have you thought about it}'' was useful because it prevented them from missing a potentially important point.
Similarly, P04 found that the partner helped reveal when they were ``\textit{leaning on just one side of the coin}'' rather than considering another point of view.


\namedparagraph{KF5: Ignoring preserved writing flow and occurred more often after pauses}
\label{sec:ignoring_flow}
Across all surfaced interventions (\tabref{tab:interaction_metrics}), the most common response was to ignore the displayed tag and continue writing without clicking to view the suggestion, accounting for 58.27\% of interventions (641/1,100).
Participants described keep typing as a natural, low-cost behavior to dismiss the interventions.
When users already knew what to write next, they simply ignored the appeared interventions.
As P13 noted, ``\textit{There are times I did not click on the little bubble thing and I was like, `No, I'm good.'}''
P05 similarly characterized ignoring as part of the natural interaction rhythm: ``\textit{if I have time I will read it, or if I think I need it I will read it, and then it just helps in a more organic way.}''
Interventions surfaced after a pause were especially likely to be ignored (64.39\%), which echos the inherent ambiguity of event triggers: a pause in typing can signal genuine cognitive need, but equally indicates focused concentration (P01).\looseness=-1

\namedparagraph{KF6: Executing was selective and depended on readiness to externalize text}
Executing represented a more committed form of engagement and accounted for 22.18\% of interventions (244/1,100), with 84.43\% (206/244) of the generated text accepted.
Interventions triggered by text selection were more likely to lead to executing (54.48\%) than those triggered by a pause (20.61\%) or sentence end (28.81\%).
Participants described letting the AI to write for them in three situations. First, they sometimes used it out of curiosity, as a low-stakes way to preview how the AI might develop a direction before deciding whether to write it themselves (P03). Second, when they already knew what they wanted to say but struggled to phrase it, execution helped translate a clear intention into fluent text (P04, P11, P12). Third, after an idea had been sufficiently developed through a short back-and-forth with a partner, execution could become the natural final step of the interaction (P11).

Participants' willingness to have the AI write for them also depended on the partner's role and the type of suggestion. For ideation-oriented partners, executing could serve as a useful shortcut once the content or direction was sufficiently settled (P06, P11). In contrast, participants were more hesitant to generate text from partners intended to provoke reflection, such as bias or rebuttal partners. P06 explained that when ``\textit{just testing an idea\dots~I'm not like rewriting anything, I'm just saying do I really need this kind of stuff? So that is where I don't really need help me write,}'' while P10 avoided execution when further development would ``\textit{deviate us too much from what\dots~we just want one nudge and move on from there.}''
Several participants further cautioned that making the AI write for users available for every suggestion could encourage over-reliance and prematurely narrow the directions they might otherwise explore (P03, P05, P11). For example, P05 deliberately avoided generating text when he instead wanted to ``\textit{guide it to write deeper}'' through further suggestions. He also argued that \textit{not} always offering the button could be beneficial because otherwise ``\textit{we are prone to just maybe always click it and actually it would limit the creativity\dots~the human part of it.}''

\begin{takeawaybox}{RQ2 Takeaway}
\textit{Participants engaged with proactive suggestions through graduated levels of attention and control.
Ignoring was the most common response and used to preserve writing flow, especially after pauses.
When support was needed, writers opened suggestions and treated them as prompts for idea generation and self-monitoring.
Executing represented a more committed form of engagement and was most appropriate when writers had a clear direction or sufficiently developed idea to externalize, particularly after selecting text as a specific target for support; participants were more hesitant to let the AI execute suggestions intended primarily to provoke reflection.}
\end{takeawaybox}

\subsection{RQ3: User Experience}
Participants' overall impressions of this new form of customizable, proactive cognitive support were positive.
In the daily diary surveys, participants gave an average rating of 8.11/10 ($SD=1.87$) for helpfulness and 7.85/10 ($SD=1.86$) for timeliness.
We next present interview findings on how participants perceived system initiative (KF7), how they thought proactive cognitive support should be timed (KF8), and how they perceived the visual (KF9) and rhetorical (KF10) representations of proactive support.\looseness=-1

\namedparagraph{KF7: Proactivity shifted support from pulling to pushing}
Participants contrasted proactive support with conventional chatbot interactions.
Interacting with chatbots required writers to leave composition, formulate a prompt, wait for a response, and then re-enter the writing task; this was especially difficult when writers did not yet know what help to request.
As P08 put it, ``\textit{sometimes it's like I don't even know what question to ask}.''
P15 described chatbot interaction as a fragmented ``\textit{spot}'' experience: ``\textit{working with Gemini, it's kind of here's a spot, there's a spot, there's a spot}.''
Proactive support, by contrast, felt more continuous because partners could surface help from within the writing context; P15 described it as ``\textit{a kind of flow and then it's a little bit like waves}.''
This shift came from the context-aware ``push mechanism'' afforded by proactivity: as P10 explained, ``\textit{it's a push mechanism rather than a pull mechanism\dots~I don't need to again click and put the information in a different thing and ask. It's sort of automated on what I want to ask since it's a partner.}''\looseness=-1

\namedparagraph{KF8: Good timing depended on contextual alignment with writers' momentary intentions}
Participants define a good timing as whether the assistance matched what they were trying to do at that particular moment. 
They did not judge timing by whether a suggestion appeared after the ``right'' interaction event but more by the alignment with their defined contextual heuristics.
The same observable event could correspond to different writing states and intentions. For example, a pause could reflect being stuck and needing help, but it could also reflect developing an idea or thinking through what to write next. 
P16 described a well-timed intervention as appearing when they were thinking ``\textit{this time I need a help},'' while P12 described pauses caused by ``\textit{brain fade or writer's block mid-sentence}'' as moments when a partner could help carry a thought forward.
Similarly, P06 distinguished sentence endings as just textual boundaries and as moments when ``\textit{I wrote a sentence, I'm done with that, now I want to move on to the next one}.''
Poor timing occurred when this alignment broke down, for example when the support did not match what the writer currently needed or when a relevant partner remain silent when they should appear.
P13 recalled one such missed opportunity: ``\textit{Oh, come on. That was a good time for one to like pop in.}'' 
\looseness=-1

\namedparagraph{KF9: Peripheral and fading visual representations made proactive support feel non-intrusive}
Participants emphasized that the visual representation of proactive suggestions shaped whether system initiative felt intrusive.
Two design features were particularly important: peripheral placement, through the right-side panel and small floating tags, and fading presence, through automatic dismissal after periods of inaction.
The right-side placement kept suggestions outside the main writing workspace and made them easy to ignore when writers wanted to stay focused.
P10 noted that suggestions ``\textit{come on the right tab and not on the main writing screen\dots~they're very small which means that they don't take up a lot of your space even on the right side.}''
P16 similarly explained, ``\textit{it's in the set panel right so it's not blocked my main\dots~if it's good then I look at that if not I just ignore.}''
The initial floating tags further reduced disruption by signaling a suggestion without immediately expanding its content.
P13 valued that ``\textit{it's just a little icon and it's not opening this whole box like out of here.}''
The fading presence of suggestions further reinforced this optionality: when writers did not engage, suggestions automatically disappeared rather than remaining in the workspace. As P12 put it, it was ``\textit{perfect because when I didn't want them, it kind of disappeared.}''\looseness=-1

\namedparagraph{KF10: Acknowledgments built trust, while questions preserved sense of control}
Participants also reflected on how the rhetorical form of proactive suggestions shaped their experience. 
The acknowledgment at the beginning of each suggestion often served a trust-building function by showing that the partner had interpreted the current writing context before offering support. P12 described this as a ``same wavelength'' signal: ``\textit{that gives me the confidence that what it's going to propose next will be relevant\dots~otherwise it could just be like a random suggestion}.'' 
P15 similarly noted that the acknowledgment prompted a useful check on whether the partner had correctly understood their intention. 
For writers with a clear sense of direction, however, this framing could feel unnecessary; P07 remarked, ``\textit{I know what I write; I just want to know what goes next}.''
Participants also generally preferred the second part of the suggestion to be framed as a question rather than a directive because questions preserved room for interpretation and choice. As P13 explained, ``\textit{it's a partner. it's not telling me what to do; it's just offering me suggestions,}'' while P15 said that questions gave ``\textit{the impression that I have still everything under control and I can decide.}''
This framing was less effective when writers lacked the knowledge needed to interpret the question. 
P08 described this as an ``unknown unknowns'' problem: ``\textit{You can tell me that I missed something, but you actually have to tell me what I missed because the reason why I didn't think about it is because I don't know about it.}''
In such cases, participants often continued the interaction by asking the partner for examples or additional options.\looseness=-1


\begin{takeawaybox}{RQ3 Takeaway}
\textit{Participants valued proactive cognitive support because it reduced the effort of explicitly seeking help and allowed assistance to emerge within the flow of writing. However, system initiative was experienced positively when the support aligned with writers' momentary intentions and remained easy to ignore or reject. Visual presentation and rhetorical framing all shaped this experience: peripheral and fading visual cues reduced disruption, acknowledgments signaled contextual understanding, and question-based suggestions preserved writers' sense of control.}
\end{takeawaybox}

%% file: tables/tab-partners.tex
\begin{table*}
\centering
\footnotesize
\caption{\formatcaption{Examples of proactive thought partners created by participants.}{The examples illustrate variation in partner roles and contextual heuristics, as well as participants' frequent use of all three event triggers. Triggers are abbreviated as P = Long Pause, E = Sentence End, and S = Text Selection.}}
\label{tab:representative_partners}
\begin{tblr}{
  width = \linewidth,
  colspec = {X[1.5,l] X[2.4,l] c X[2.2,l]},
  hline{1,Z} = {0.8pt, solid},
  row{1} = {font=\bfseries},
  hline{2} = {0.5pt, solid},
  row{1} = {bg=gray!10},
  cell{2-7}{1} = {r=1}{font=\bfseries,bg=gray!6},
}
Partners & Roles & Triggers & Heuristics \\

Research Partner
& {\taglabel{Information Seeking}
{Provide examples of similar research topics and findings to augment my essay.}}
& P / E / S
& {\taglabel{Writing-Activity \& Anticipated-Need} {When I'm trying to quote examples and trying to back it up with broader facts.}} \\

India Partner
& {\taglabel{Information Seeking} {Bring up India-centric market scenarios that drive pricing, including behavior of the Indian customer across industries.}}
& P / E / S
& {\taglabel{Draft-State \& Writing-Activity} {When I talk about India as a country and drill down into specific scenarios.}} \\

Internal Monologue Partner
& {\taglabel{Ideation Support}
{Provides quotes for what could be running through a character's head at certain moments within the story.}}
& P / E / S
& {\taglabel{Anticipated-Need}
{When it would be helpful to know what the protagonist is thinking.}} \\

Ethical partner
& {\taglabel{Critical Reflection}
{You're a senior ethical program manager. Defend the ethical boundaries of ideas without losing sights of values.}}
& P / E / S
& {\taglabel{Draft-State}
{When there are cases of challenging ethical issues when using AI.}} \\

Synthesis Partner
& {\taglabel{Argument Development}
{Help integrate positive and negative arguments into a coherent conclusion.}}
& P / E / S
& {\taglabel{Draft-State \& Writing-Activity \& Anticipated-Need}
{When I have presented both sides of the argument, or will soon need to unify my points.}} \\

Corporate Speak Partner
& {\taglabel{Textual Assistance}
{Up-level the language to make it more professional and integrate business speak.}}
& E
& {\taglabel{Draft-State}
{When my language is too casual and does not reference business talk.}} \\

\end{tblr}
\end{table*}

%% file: tables/tab-engagement.tex
\begin{table}
\centering
\footnotesize
\caption{\formatcaption{Distribution of engagement forms across trigger events.}{Each row describes interventions surfaced by a trigger event and how participants engaged with them: whether they ignored the intervention, used it as inspiration, or asked the partner to execute follow-up changes. Because Executing required first opening the suggestion, Executing cases are coded only under Executing rather than double-counted under Inspiring. Percentages are calculated within each row.}}
\label{tab:interaction_metrics}
\begin{tabular}{l r@{\ }l r@{\ }l r@{\ }l r}
\toprule
\multirow{2}{*}{\textbf{Trigger Events}}
& \multicolumn{6}{c}{\textbf{Engagement Forms}}
& \multirow{2}{*}{\textbf{Total}} \\
\cmidrule(lr){2-7}
& \multicolumn{2}{c}{\textbf{Ignoring}} 
& \multicolumn{2}{c}{\textbf{Inspiring}} 
& \multicolumn{2}{c}{\textbf{Executing}} 
& \\
\midrule
\textbf{Pause} 
& 425 & (64.39\%) 
& 155  & (23.48\%) 
& 80 & (12.12\%) 
& 660 \\

\textbf{Sentence} 
& 165 & (55.93\%) 
& 45  & (15.25\%) 
& 85  & (28.81\%) 
& 295 \\

\textbf{Selection} 
& 51  & (35.17\%) 
& 15  & (10.34\%) 
& 79  & (54.48\%) 
& 145 \\

\textbf{Total}
& 641 & (58.27\%)
& 215 & (19.55\%)
& 244 & (22.18\%)
& 1,100 \\

\bottomrule
\end{tabular}
\end{table}

%% file: sections/7-discussion.tex
\section{Discussion}
In this paper, we explored proactive thought partners, a design concept that combines system initiative with customizable, cognitive support for writing.
Drawing on findings from our probe study, we discuss design implications for future proactive writing support across four dimensions: customization, timing, engagement, and representation (\tabref{tab:design_implications}).\looseness=-1

\input{tables/tab-designimplications}

\subsection{Designing Customization as Prospective Planning}

Participants used customization to envision how proactive support could fit an upcoming writing task (\textbf{KF1}). 
Configuration externalized a situated model of forthcoming writing work: the goals they wanted to achieve, the difficulties they anticipated, and the kinds of cognitive support that might become useful.
This extends prior work on customizable AI agents for writing feedback~\citep{benharrakWriterDefinedAIPersonas2024a} after writing.
In proactive writing contexts, however, writers needs to anticipate support before and/or during writing. 
Future systems should therefore support this prospective planning by helping writers translate anticipated goals and difficulties into specialized partners for different forms of cognitive work, such as information seeking, argument development, critical reflection, and ideation (\textbf{DI1--DI2}).
Systems could help writers assemble a set of partners whose expertise and responsibilities reflect the demands of a particular writing task.

At the same time, prospective planning is necessarily imperfect because writers may not know in advance which configurations will be useful once writing unfolds. Future systems should therefore treat partner configuration as an iterative design activity instead of a one-time setup step. Writers could preview a partner's likely interventions on a sample passage, simulate how its role and timing conditions would behave in different writing situations, or revise configurations based on use. For example, a system might surface patterns such as a partner being repeatedly ignored, rarely activated, or frequently invoked for follow-up, and use these traces to suggest refinements to its role or scope. Such mechanisms could make proactive partners easier to test and adapt while preserving writers' authority over what kinds of cognitive support they want in their writing process.

\subsection{Designing Timing as Contextual Alignment}
Our findings show that observable interaction events should be treated as candidate moments for proactive support, not as sufficient evidence that an intervention is needed (\textbf{DI3}). Participants frequently configured pauses, sentence endings, and text selections as triggers, yet the same event could correspond to different writing states (\textbf{KF3}, \textbf{KF8}). A pause might indicate writer's block, focused reflection, or a moment when the writer already knows what to write next; a sentence ending might signal completion, extension, verification, or reconsideration. This distinction complements prior work on interruption management~\citep{baileyEffectsInterruptionsTask2001,czerwinskiInstantMessagingEffects2000,iqbalIndexOpportunity2005} and complicates proactive and autonomous writing systems that tie system action directly to predefined events~\citep{lehmannCollaborativeDocumentEditing2026,puAssistanceDisruptionExploring2025a}. For higher-level cognitive support, an event can indicate when the system should assess the situation, but it does not by itself determine whether support is appropriate.

Future proactive writing systems should therefore add a layer of contextual inference between event detection and intervention (\textbf{DI4}). Our participants' contextual heuristics drew on draft-state cues, ongoing writing activities, and anticipated support needs, suggesting that systems should interpret interaction events in relation to what is currently in the draft, what the writer appears to be doing, and what kind of help may be useful next. Behavioral events such as pauses and selections remain informative, but their meaning is shaped by the recursive interplay of planning, translating, and reviewing in writing~\citep{flowerCognitiveProcessTheory1981a}. Systems could use triggers to initiate contextual assessment, then combine recent writing behavior, local and document-level content, and writer-defined goals to decide whether and which partner should intervene. They should also reassess this interpretation close to delivery time, since a need inferred at one moment may no longer be relevant after the writer has moved on.


\subsection{Designing Engagement as Graduated Commitment}

Participants engaged with proactive suggestions through distinct levels of attention and commitment: ignoring them, opening them for inspiration or reflection, and selectively asking partners to execute changes directly in the document. This finding speaks to recent concerns about agency and ownership in AI co-writing. Prior work has shown that AI-generated text can complicate writers' perceptions of authorship~\citep{draxlerAIGhostwriterEffect2024}, while interaction designs can preserve agency by giving writers meaningful control over how AI contributions enter a draft~\citep{carreraWhereAddEgg2025,yehGhostWriterAugmentingCollaborative2025}. In our study, ignoring was the most common response and functioned as a natural way to preserve writing flow when writers did not need help (\textbf{KF5}). Future systems should therefore make non-engagement an effortless and legitimate response to system initiative, allowing writers to continue composing without requiring explicit dismissal (\textbf{DI5}). When support was useful, participants opened suggestions as prompts for idea generation and self-monitoring (\textbf{KF4}), suggesting that proactive assistance should first offer low-commitment material that writers can inspect, question, or adapt without immediately changing the draft (\textbf{DI6}).\looseness=-1

Direct execution represented a deeper level of commitment and was most appropriate when writers were ready to externalize a sufficiently settled intention, such as when they had a clear direction or target for revision, or had developed an idea through interaction with a partner (\textbf{KF6}). At the same time, participants often preferred not to execute suggestions intended primarily to provoke reflection or exploration, where the suggestion itself could be sufficient. This suggests that proactive systems should separate the initiation of support from the authority to modify the document and treat execution as a selective escalation (\textbf{DI7}). A system may proactively surface an idea, question, or possible revision, while direct insertion or rewriting should remain a deliberate user action. 
Future interfaces could support smooth escalation across these levels, letting proactive AI take initiative without assuming control over whether or how its contributions ultimately enter the draft.\looseness=-1

\subsection{Designing Representation as Lightweight and Non-directive}

Participants valued proactive support when its visual and rhetorical representation made system initiative feel optional and easy to assess (\textbf{KF9--KF10}). Peripheral placement, small floating tags, and fading presence allowed suggestions to remain available without competing with the main writing task, suggesting that proactive systems should present interventions as lightweight bids for attention that can recede when writers choose not to engage (\textbf{DI8}). Rhetorical framing also shaped how writers interpreted these interventions. Acknowledgments helped establish trust by making the system's understanding of the local context visible, while non-directive suggestions (\ie questioning) preserved room for interpretation and choice (\textbf{DI9}). Future systems should therefore communicate enough of their contextual interpretation for writers to judge whether a suggestion is relevant, while avoiding forms that make the intervention feel directive or demanding.

This finding aligns with work on interruption management showing that interruption costs depend not only on timing but also on how an intervention is presented~\citep{baileyMeasuringEffectsInterruptions2000,czerwinskiInstantMessagingEffects2000,puAssistanceDisruptionExploring2025a}.
Future representations could make this grounding visible at multiple levels of the document. Local interventions might reference the sentence or passage that motivated a suggestion, while partners concerned with argument flow, section coherence, or narrative direction could surface their interpretation of broader document structure through outlines, highlighted gaps, or cross-section references. Such representations could help writers understand why a proactive suggestion appeared while keeping the intervention lightweight and subordinate to the writing task.





\section{Ethical Considerations}
Proactive writing support raises privacy concerns because providing timely assistance requires continuous observation of both the evolving document and the writer's interaction traces. 
In our probe, the system used the current draft, cursor position, behavioral events, and recent keystroke logs to infer when support might be needed.  
Such data may expose sensitive content as well as distinctive writing habits that could reveal aspects of a writer's identity. 
Although we instructed participants not to enter personally identifiable, confidential, or highly sensitive information, this mitigation places responsibility on users and may be unrealistic in everyday writing contexts. 
Future systems should therefore adopt privacy-by-design practices~\citep{spiekermannChallengesPrivacyDesign2012}, including data minimization, local or ephemeral processing of behavioral traces, strict retention limits, clear disclosure of what is monitored, and granular controls for pausing monitoring or deleting collected data. 
Proactive assistance should also avoid treating inferred cognitive states as definitive and should not reuse writing traces for model training, profiling, or identity inference without separate, explicit consent.

\section{Limitations}
Our study has several limitations.
First, as a probe study, it was intended to explore an emerging design space rather than evaluate a finished system.
Accordingly, our findings do not establish causal effects on writing quality, productivity, learning, or long-term agency. 
Longer deployments or a larger scale of comparative, quantitative lab studies against reactive AI assistants and conventional writing tools can be valuable in the future.
A related limitation is that our study can only offer preliminary insight into how proactive support may influence writers' sense of ownership. 
Although participants retained control over whether to ignore, check, or execute suggestions, proactive interventions may still shape what writers notice, how they frame problems, and which directions they pursue. 
These effects may become more consequential over longer periods of use, especially if writers begin to rely on partners to identify writing problems or propose next steps. 
Future work could examine how proactive support affects perceived authorship, confidence, skill development, and dependence over time.\looseness=-1

Our participant sample may limit generalizability.
Participants were English-proficient daily writers who engaged with a range of writing genres during the study, and all had prior experience with AI writing tools. 
This background allowed them to compare proactive support with existing AI writing tools (\eg chatbots, autocomplete) during interviews, but their expectations, comfort with AI, and willingness to engage with proactive support may differ from those of writers with less AI experience or different linguistic backgrounds.
In addition, our privacy instruction to avoid confidential workplace content likely excluded the high-stakes writing these participants do most.
Future work could address these limitations by evaluating proactive thought partners with a more demographically and linguistically diverse participant population, including writers with varying levels of AI experience, and across a broader range of professional and high-stakes writing contexts.\looseness=-1

Our probe also has several limitations. First, current LLMs may not reliably interpret writing context well enough to infer writers' momentary intentions or anticipated support needs. Our findings around contextual heuristics provide a starting point for future work by identifying three forms of contextual cues that may be useful for this inference: draft-state cues, writing-activity cues, and anticipated-need cues. Future research could develop models and benchmarks that evaluate whether AI systems can recognize these cues, distinguish among different writing states that produce similar interaction events, and determine which forms of support are appropriate in context. Second, suggestion quality and timing depended on a particular LLM-based implementation. Different models, prompts, contextual representations, or orchestration strategies may produce different intervention patterns and user experiences. 
Our findings should therefore be interpreted as reflecting both the broader interaction paradigm of proactive thought partners and the specific capabilities and limitations of the system we implemented.\looseness=-1

%% file: tables/tab-designimplications.tex
\begin{table*}
\centering
\footnotesize
\caption{\formatcaption{Design dimensions, factors, and implications.}{This table summarizes the design dimensions of proactive thought partners explored through our probe, the design factors examined in the study, and the resulting implications for future customizable proactive writing support.}}
\label{tab:design_implications}
\begin{tblr}{
width = \linewidth,
colspec = {Q[0.17\linewidth,l] X[1.2,l] X[2.15,l]},
hline{1,Z} = {0.8pt, solid},
hline{4,6,9} = {0.4pt, solid},
row{1} = {font=\bfseries},
row{1} = {bg=gray!10},
cell{2}{1} = {r=2}{font=\bfseries,bg=gray!6},
cell{4}{1} = {r=2}{font=\bfseries,bg=gray!6},
cell{6}{1} = {r=3}{font=\bfseries,bg=gray!6},
cell{9}{1} = {r=2}{font=\bfseries,bg=gray!6},
rowsep = 3pt,
colsep = 5pt,
}
Design Dimension & Design Factors & Design Implications \\
\hline

\textbf{Customization}
& \textbf{Configuration Strategies:} goal-driven or difficulty-driven planning.
& \textbf{[DI1]} Support prospective planning by helping writers anticipate the goals, difficulties, and forms of assistance relevant to an upcoming writing task, and translate these expectations into partner configurations. \\

& \textbf{Partner Functionality:} information seeking, argument development, critical reflection, and ideation support.
& \textbf{[DI2]} Support specialized partners for distinct forms of cognitive work rather than defaulting to general-purpose assistants. Systems should make it easy to assemble partners around the knowledge, reasoning, reflection, or ideation support relevant to a particular writing task. \\

\textbf{Timing}
& \textbf{Event Triggers:} long pause, sentence end, and text selection.
& \textbf{[DI3]} Treat observable interaction events as candidate moments for intervention rather than sufficient evidence of need. The same event may reflect different writing states, so triggers should initiate further assessment rather than directly determine system action. \\

& \textbf{Contextual Cues:} draft-state cues, writing-activity cues, and anticipated-need cues.
& \textbf{[DI4]} Infer intervention opportunities from multiple contextual cues that capture what is present in the draft, what the writer is currently doing, and what support they may need next. Systems should align assistance with the writer's momentary intention before deciding whether and which support to provide. \\

\textbf{Engagement}
& \textbf{Ignoring:} users can naturally ignore proactive interventions by continuing to write.
& \textbf{[DI5]} Make ignoring a first-class response to system initiative. Proactive suggestions should be easy to bypass without requiring explicit dismissal or disrupting the writing flow. \\

& \textbf{Inspiring:} users can inspect suggestions and optionally continue a dialogue.
& \textbf{[DI6]} Provide low-commitment ways to use proactive suggestions as material for thinking. Writers should be able to inspect, question, elaborate, or adapt a suggestion without immediately incorporating AI-generated text into the draft. \\

& \textbf{Executing:} writers can ask partners to insert or revise text directly.
& \textbf{[DI7]} Reserve direct text generation for moments when writers are ready to externalize a sufficiently settled intention. Systems should allow writers to escalate from cognitive support to execution when they have a clear direction or target for revision, while avoiding making execution the default for suggestions intended to provoke reflection or exploration. \\

\textbf{Representation}
& \textbf{Visual Form:} peripheral placement and fading presence.
& \textbf{[DI8]} Represent proactive suggestions as lightweight, peripheral bids for attention. Suggestions should remain noticeable without occupying the primary writing space and should recede when writers choose not to engage. \\

& \textbf{Suggestion Framing:} acknowledgment of writer intention and thought-provoking questions.
& \textbf{[DI9]} Make the system's contextual interpretation visible while preserving room for writer judgment. Brief acknowledgments can communicate what the system believes the writer is doing, while question-based suggestions can invite reflection without prescribing a particular response. \\

\end{tblr}
\end{table*}

%% file: sections/8-conclusion.tex
\section{Conclusion}

We introduced proactive thought partners, a design concept for AI writing assistants that proactively provide customizable, higher-level cognitive support during writing. Through a one-week deployment of our technology probe, we found that writers used customization as a form of prospective planning, engaged with proactive suggestions through graduated levels of commitment, and experienced good timing as contextual alignment between offered support and their momentary intentions.
Our findings further show that effective proactivity depends on more than deciding when to intervene: writers need ways to shape the kinds of support they receive, ignore or deepen engagement with suggestions as needed, and receive proactive support through non-intrusive visual and rhetorical representations.
We believe our design and insights can inform future proactive AI systems for complex cognitive work.\looseness=-1